\documentclass[journal=nalefd,manuscript=letter]{achemso}
\usepackage{achemso}

\usepackage{float} 
\usepackage{subfigure}
\usepackage{graphicx}%
\usepackage{hyperref}
\usepackage{dcolumn}
\usepackage{bm}
\usepackage{soul} 
\usepackage{xcolor}
\usepackage{amsmath}

\title{Non-Invasive Readout of the Kinetic Inductance of Superconducting Nanostructures}

\author{Lukas Nulens}
\email{lukas.nulens@kuleuven.be}
\affiliation{Quantum Solid-State Physics, Department of Physics and Astronomy, KU Leuven, Celestijnenlaan 200D, B-3001 Leuven, Belgium}

\author{Davi A. D. Chaves}
\affiliation{Departamento de Física, Universidade Federal de São Carlos, 13565-905 São Carlos, SP, Brazil}
\author{Omar J. Y. Harb}
\affiliation{Quantum Solid-State Physics, Department of Physics and Astronomy, KU Leuven, Celestijnenlaan 200D, B-3001 Leuven, Belgium}

\author{Jeroen E. Scheerder}
\affiliation{Imec, Kapeldreef 75, 3001 Leuven, Belgium}
\author{Nicolas Lejeune}
\affiliation{Experimental Physics of Nanostructured Materials, Department of Physics , Université de Liège, Allée du 6 Août 19, B-4000 Sart Tilman, Belgium}

\author{Kamal Brahim}
\affiliation{Imec, Kapeldreef 75, 3001 Leuven, Belgium}

\author{Bart Raes}
\affiliation{Imec, Kapeldreef 75, 3001 Leuven, Belgium}
\author{Alejandro V. Silhanek}
\affiliation{Experimental Physics of Nanostructured Materials, Department of Physics , Université de Liège, Allée du 6 Août 19, B-4000 Sart Tilman, Belgium}

\author{Margriet J. Van Bael}
\affiliation{Quantum Solid-State Physics, Department of Physics and Astronomy, KU Leuven, Celestijnenlaan 200D, B-3001 Leuven, Belgium}

\author{Joris Van de Vondel}
    \email{joris.vandevondel@kuleuven.be}
\affiliation{Quantum Solid-State Physics, Department of Physics and Astronomy, KU Leuven, Celestijnenlaan 200D, B-3001 Leuven, Belgium}

\begin{document}

\begin{abstract}
The energy landscape of multiply connected superconducting structures is ruled by fluxoid quantization due to the implied single-valuedness of the complex wave function. The transitions and interaction between these energy states, each defined by a specific phase winding number, are governed by classical and/or quantum phase slips. Understanding these events requires the ability to probe, non-invasively, the state of the ring. Here, we employ a niobium resonator to examine the superconducting properties of an aluminum loop. By applying a magnetic field, adjusting temperature, and altering the loop's dimensions via focused ion beam milling, we correlate resonance frequency shifts with changes in the loop's kinetic inductance. This parameter is a unique indicator of the superconducting condensate's state, facilitating the detection of phase slips in nanodevices and providing insights into their dynamics. Our method presents a proof-of-principle spectroscopic technique with promising potential for investigating the Cooper pair density in inductively coupled superconducting nanostructures.

% \lukas{Original}\\
% The energy landscape of superconducting ring-like structures is ruled by quantization of fluxoid due to the implied single-valuedness of the complex wave function. The transitions and interaction between these states, each characterized by a particular winding number of the superconducting phase, are completely determined by the occurrence of classical and/or quantum phase slips. Understanding these events requires the ability to probe, non-invasively, the state of the ring. Here, we utilize a niobium resonator as a non-invasive probe to explore the superconducting properties of an aluminum loop.  By locally applying a magnetic field, adjusting the temperature, and modifying the loop dimensions using focused ion beam milling, we can directly link the shifts in the resonance frequency to changes induced in the kinetic inductance of the aluminum ring. The kinetic inductance is a unique indicator characterizing a particular state (winding number) of the superconducting condensate. Therefore, it provides an effective parameter for detecting phase slips in nanodevices, offering insights into the dynamics of these quantum phenomena. The proposed approach depicts a proof of principle spectroscopy technique showing promising perspectives for exploring the Cooper pair density within inductively coupled superconducting nanostructures. 
\end{abstract}

\section{Introduction}
\label{sec:intro}
Understanding phase slips, topological fluctuations of the superconducting order parameter in superconducting loops, is crucial for the development of future nanodevices. These phenomena are foundational for certain superconducting qubits \cite{mooij2005phase,mooij2006superconducting}, realized in disordered superconductors like InO$_x$\cite{astafiev2012coherent} and NbN \cite{peltonen2013coherent,de2018charge}. Additionally, phase-slip-based devices have applications in memory elements\cite{ligato2021preliminary,chaves2023nanobridge,murphy2017nanoscale}, charge quantum interference devices \cite{de2018charge}, and as a possible quantum current standard \cite{shaikhaidarov2022quantized}.

Fluxoid quantization \cite{london1950superfluids, tinkham2004introduction} is evidenced by the analysis of a superconducting ring subjected to a magnetic field. This quantization arises from the single-valuedness of the wavefunction, implying that its phase can only change in integer multiples of $2\pi$. This leads to distinct states in the ring, each associated with a specific winding of the phase denoted by its winding number, which can be altered by the occurrence of a phase slip event. Each state corresponds at a particular magnetic field value to a unique amount of circulating current, directly influencing the condensate’s strength by reducing the Cooper pair density $n_s$. As the kinetic inductance $L_k$ is inversely proportional to $n_s$, a high-resolution $L_k$ probe should be able to distinguish states and transitions.

Even though studies have detected phase slips in DC measurements on superconducting loops \cite{murphy2017asymmetric,nulens2022metastable}, measuring phase slips and determining  $L_k$ in DC experiments remains challenging \cite{baumans2016thermal,baumans2017statistics,massarotti2015breakdown,tafuri2019fundamentals,aref2012quantitative}. Consequently, embedding superconducting quantum interference devices (SQUIDs) and loop-like structures in gigahertz range resonator systems \cite{belkin2015formation, kennedy2019tunable, segev2011metastability, khabipov2022superconducting,potter2023controllable} and alternative strategies based on scanning probe techniques \cite{polshyn2018imaging}, cantilever torque magnetometry \cite{petkovic2016deterministic,jang2011observation}, and SQUID magnetometry \cite{zhang1997susceptibility,silver1967quantum} have helped to better understand the phase slip dynamics.

Here we introduce an innovative approach for measuring the kinetic inductance of a superconducting aluminum loop using high-frequency spectroscopy. Rather than physically connecting a loop to a resonator \cite{astafiev2012coherent,peltonen2013coherent,de2018charge,skacel2019development}, we position it in the gap of a niobium coplanar waveguide resonator. A flux bias line is used for the local application of a magnetic field at the Al loop. This architecture allows us to modify
the loop's properties through adjustments in the magnetic field or
temperature while leaving the Nb readout resonator unaffected \cite{nulens2023catastrophic}. The field and temperature dependencies of the resonance frequency $f_r(B,T)$ can be directly linked to changes in the kinetic inductance and Cooper pair density of the Al ring. As such, it is possible to probe $L_k$ of the Al ring without ever leaving the superconducting state, a crucial requirement for the functionality of non-volatile memory elements \cite{ligato2021preliminary}. For each state, a unique magnetic field dependence of the resonance frequency is obtained, where various observed discontinuities mark the sudden changes in the winding number of the loop due to phase slips. As the kinetic inductance of the structure is increased by altering the width of one section of the loop through focused ion beam (FIB) milling, an earlier occurrence of the initial phase slip indicates a successful reduction in the characteristic energy barrier preventing these events. The strong correlation between the readout resonator's behavior and the loop's kinetic inductance demonstrates the potential of this spectroscopy technique for sensitive probing of inductively coupled superconducting nanostructures and sets the stage for phase slip manipulation and detection.

\section{Working principle}
\label{sec:design}

\begin{figure}[H]
    \centering
    \includegraphics[width=\columnwidth]{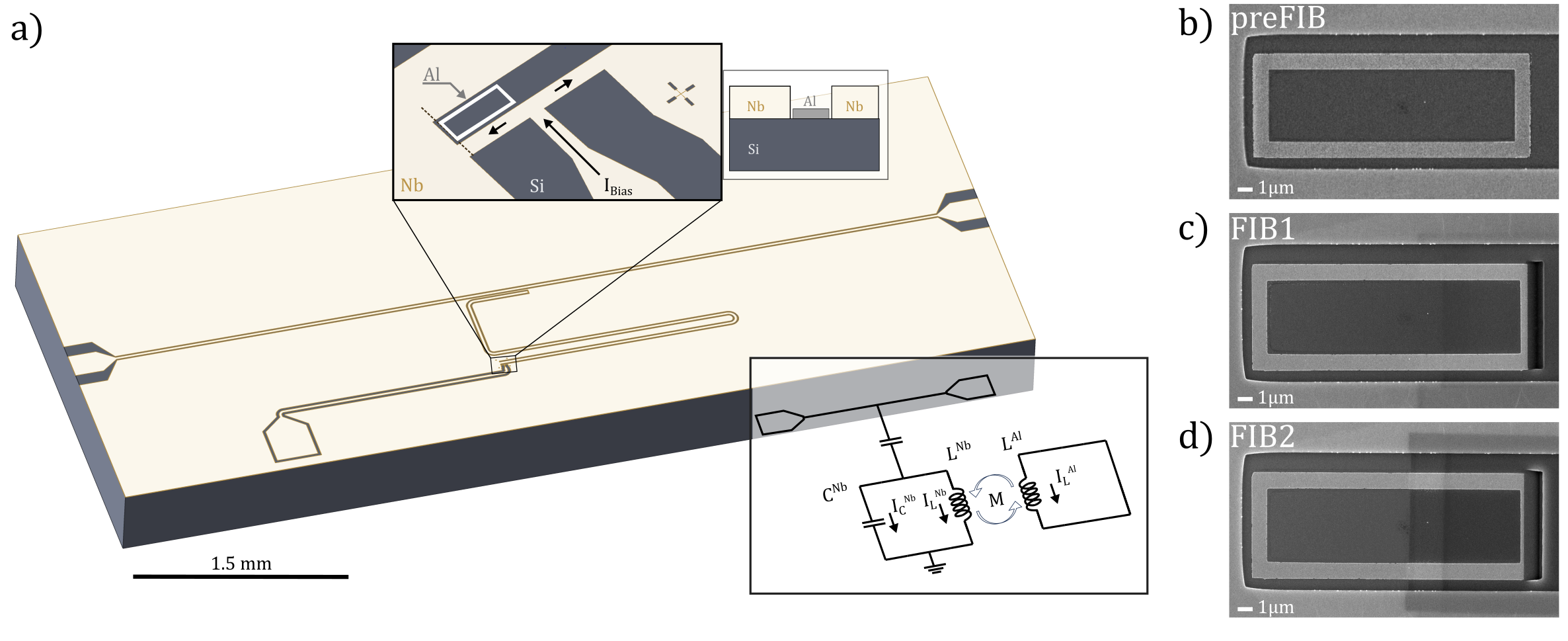}
    \caption{Device layout and electrical equivalent. a) A schematic representation of the device under investigation, comprised of a $\lambda/4$ resonator capacitively coupled to the central feedline. Niobium is represented in white while the silicon substrate is represented in dark gray. A flux bias line is connected to the resonator's end piece to supply a local magnetic field in the vicinity of an Al loop, positioned at the Si trench and evident in the close-up view in the upper inset. The accompanying side view corresponds to a cross section along the dashed line at the edge of the Al loop. The circuit diagram representation of the resonator-loop system is provided in the bottom inset. Scanning electron microscope (SEM) image of the aluminum loop b) before any alterations in its dimensions (preFIB), c) after the reduction of one arm of the loop to 500~nm (FIB1), and d) after a further reduction down to 200~nm (FIB2).}
    \label{fig:fig1_sample}
\end{figure}

The resonator consists of a 100-nm-thick Nb layer deposited on a 525~µm high-resistivity silicon substrate ($\rho > 20~\text{k}\Omega \text{cm}$). A local flux bias line was designed to mitigate the complex field penetration into the coplanar waveguide resonator subjected to a perpendicular magnetic field,\cite{nulens2023catastrophic} as visualized in Figure \ref{fig:fig1_sample}a. The device comprises a hanger-type $\lambda/4$ coplanar waveguide resonator with a length of 4561 µm, capacitively coupled to a central feedline with a width $w =$ 20 µm and a gap of 10 µm on a $(7 \times 3)$ mm$^2$ substrate. The high-frequency (4--8 GHz) measurements were performed using a Keysight P5003B Streamline Vector Network Analyzer at a driving power of 0 dBm in a Janis $^3$He cryostat. The input line was attenuated by 60 dB distributed across different temperature stages within the cryostat. Upon analysis of the transmission parameter $S_{21}$ at a temperature $T$ = 0.3 K, a resonance frequency $f_r$ = 6.4755 GHz and a loaded quality factor of $Q_l$ = 3006 were obtained for the bare resonator. Then, an Al loop with inner dimensions of $(20 \times 6)$~µm$^2$, $w = 1$~µm, and thickness 50~nm was deposited on the resonator structure. The loop is located where the current through the resonator, and as such the magnetic field, is maximum \cite{pozar2011microwave,muller2013detection}, as illustrated in Figure \ref{fig:fig1_sample}a.
A $T_c$ $\sim 1.23$ K, a mean free path $l \sim$ 16 nm, and a coherence length $\xi(0) = 134$ nm were obtained from a reference Al bridge yielding an effective penetration depth of $\lambda_{eff}(0) = 196$ nm \cite{tinkham2004introduction,de2018superconductivity,pearl1964current}.
          
The resonator-loop system can be represented by an electrical equivalent circuit as depicted in Figure \ref{fig:fig1_sample}a. The Al loop is represented as an inductor $L^{\text{Al}}$ coupled through a mutual inductance $M$ to the Nb resonator. The latter is considered as a parallel inductor-capacitor circuit $L^{Nb}$- $C^{Nb}$, with a bare resonance frequency $\sim\frac{1}{\sqrt{L^{\text{Nb}}C^{\text{Nb}}}}$. The mutual inductance is defined as $M=k\sqrt{L_g^{\text{Nb}}L_g^{\text{Al}}}$, where the coupling factor $k$ lies within the interval $[0,1]$, and $L_g$ is the geometric contribution to the total inductance $L^x(T) = L^x_g + L^x_k(T)$ with $x$ equal to Nb or Al. Following Kirchhoff's laws, the equations representing the circuit can be written as
    \begin{equation*}
        \begin{bmatrix}
        	\frac{1}{j\omega C^{\text{Nb}}} & 0 & 0 \\
        	0 & -j\omega L^{\text{Nb}} & -Mj\omega \\
        	0 & Mj\omega & j\omega L^{\text{Al}} \\	
        \end{bmatrix}
        \cdot
        \begin{bmatrix}
        	I_{C^{\text{Nb}}}\\
        	I_{L^{\text{Nb}}} \\
        	I_{L^{Al}}
        \end{bmatrix}
         = 
        \begin{bmatrix}
         	V\\
         	V \\
         	0
        \end{bmatrix}
        .
        \end{equation*}
Here, $\omega$ is defined as the driving angular frequency, with $I_{C^{\text{Nb}}}$, $I_{L^{\text{Nb}}}$, and $I_{L^{\text{Al}}}$ representing the currents flowing through the different circuit elements. Following Cramer's rule for $3\times 3$ matrices and the resonance condition $I_{C^{\text{Nb}}}=I_{L^{\text{Nb}}}$ leads to a resonance frequency
    \begin{equation}
        \label{eq:fr_model}
        f_r = \frac{1}{\sqrt{16C^{\text{Nb}}L^{\text{Nb}}\left(1-k^2\dfrac{L^{\text{Nb}}_gL^{\text{Al}}_g}{L^{\text{Al}}L^{\text{Nb}}}\right)}} .
    \end{equation}
    
In addition, the kinetic inductance of a superconducting wire can be determined by equating the total kinetic energy of the Cooper pairs with an equivalent inductive energy, as described in Ref. \citenum{annunziata2010tunable}. This leads to:
     \begin{equation}
       \label{eq:Lk_BTdim}
       L_k = \frac{m^*}{4e^2}\frac{1}{n_s(T,B)}\frac{l}{wd}.
    \end{equation}
Here, $m^*$ represents the mass of a Cooper pair, $e$ the elementary charge, and $l$, $w$, and $d$ stand for the length, width, and thickness of the superconducting wire, respectively.
From Eqs. \eqref{eq:fr_model} and \eqref{eq:Lk_BTdim}, any temperature or field-induced change in the loop's $n_s$ will lead to a change in the resonator's $f_r$ by virtue of the high-quality factor achievable by superconducting resonators \cite{megrant2012planar}. This results in a very high detection sensitivity in a bandwidth limited by $f_r$/$Q_l$. However, correlating $L_k$ with geometry without fabricating entirely new devices poses a challenge. To overcome this issue, we utilize FIB milling permitting to selectively reduce one arm's dimensions, as depicted in Figure \ref{fig:fig1_sample}b, c, and d. Two FIB sessions were conducted, decreasing the width of the right-most edge of the loop from 1 µm (preFIB) to 500 nm in FIB1, and down to 200 nm in FIB2. FIB milling aims to exclusively modify a single parameter, the kinetic inductance, by reducing the width of the constriction (Eq. \eqref{eq:Lk_BTdim}). Throughout the FIB milling, the sample stayed wire-bonded, enhancing comparison robustness between sessions and minimizing potential influences beyond the change in geometry.

\section{Temperature dependence}
\label{sec:Tdep}
\begin{figure}[H]
    \centering
    \includegraphics[width=\columnwidth]{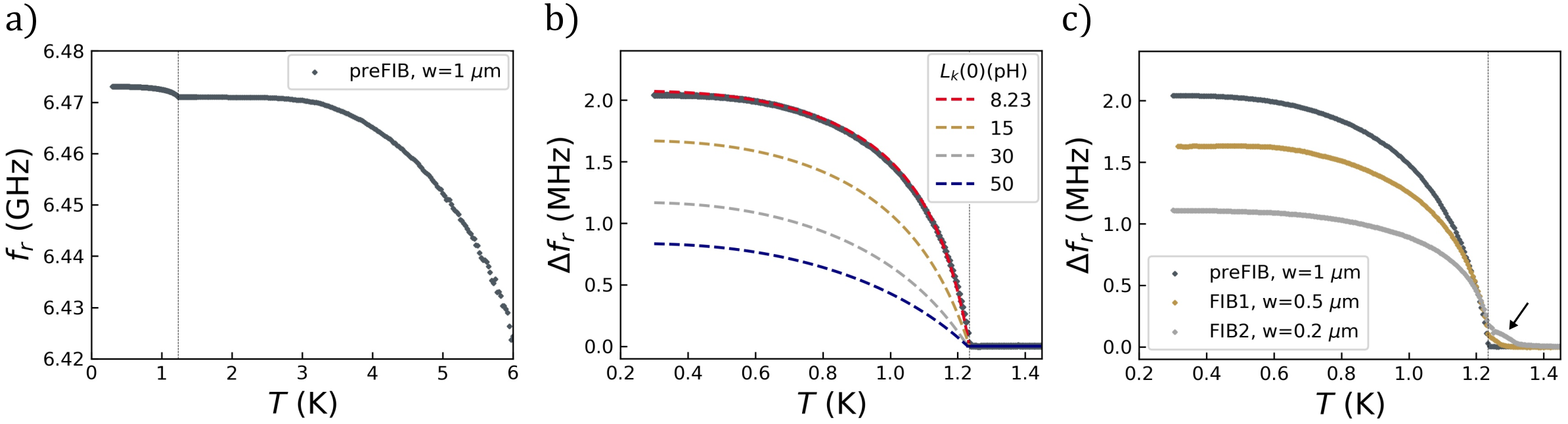}
    \caption{Temperature-dependent experimental data and model analysis. a) Temperature dependence of the Nb resonator's resonance frequency ($f_r(T)$) obtained for the preFIB sample. The critical temperature of the Al loop is indicated in all panels by a vertical dashed line. b) $\Delta f_r(T)$ obtained following Equations \eqref{eq:fr_model} and \eqref{eq:eq_Lk} for increasing values of $L^{Al}_k(0)$ as indicated in the legend and parameters $k=0.031, L^{Nb}_g=1945 \text{ pH}, C^{Nb} =0.76 \text{ pF},\text{ and } L^{Al}_g =20 \text{ pH}$. The top red dashed line corresponds to a fit of the experimental preFIB data of panel a) represented by grey markers. c) Details of $\Delta f_r(T)$ below 1.45~K before and after the FIB sessions.  The black arrow indicates the tail appearing after the FIB session related to the morphology of Al.}
    \label{fig:fig2_temp}
\end{figure}

Figure \ref{fig:fig2_temp}a shows the preFIB temperature dependence of the resonator's resonance frequency, determined using the fitting routine outlined in Ref. \citenum{probst2015efficient}. The temperature dependence of the kinetic inductance in Eq. \eqref{eq:Lk_BTdim} can be written as \cite{annunziata2010tunable,tinkham2004introduction}

\begin{equation}
    \label{eq:eq_Lk}
    L_k(T) = L_k(0)\dfrac{1}{\tanh\left(1.74\sqrt{\dfrac{T_c}{T}-1}\right)}\dfrac{1}{\tanh\left(\dfrac{\Delta(T)}{2k_BT}\right)} , 
\end{equation}
where $L_k(0) \propto \dfrac{l}{w}\dfrac{Rh}{2\pi^2\Delta(0)}$ with $l$ and $w$ corresponding to the length and width of the resonator, $R$ the normal state sheet resistance, and $h$ and $k_B$ the Planck’s and Boltzmann’s constants, respectively. The superconducting energy gap $\Delta(T)$ can be estimated by the interpolation formula $\Delta(0)\tanh(1.74\sqrt{T_c/T-1})$ \cite{nulens2023catastrophic}. In all panels of Figure \ref{fig:fig2_temp} the critical temperature of Al is indicated by a grey vertical dashed line. In Figure \ref{fig:fig2_temp}a, for temperatures exceeding $1.23$ K, the experimental data aligns with the resonance frequency of a bare Nb resonator determined by the change in its kinetic inductance as previously shown in several works \cite{barends2008contribution,frasca2023nbn,nulens2023catastrophic,annunziata2010tunable}.

As the loop transitions into the superconducting state ($T < 1.23$ K), an increase in the resonance frequency with decreasing temperature becomes apparent. To quantify this change we introduce the parameter $\Delta{f_r}$ as the difference in resonance frequency relative to its value at 1.45 K, where Al is in the normal state ($\Delta f_r(T) = f_r(T) - f_r(1.45 \ \text{K})$). In Figure \ref{fig:fig2_temp}b, the top red dashed line represents the variation of $\Delta{f_r}$ with temperature. The fit is obtained using Eqs.~\eqref{eq:fr_model} and \eqref{eq:eq_Lk}, with fixed parameters $L^{Nb}_g = 1945 \ \text{pH}$, $C^{Nb} = 0.76 \ \text{pF}$, $L^{Al}_g = 20 \ \text{pH}$, $T_c = 1.23\ \text{K}$  and fitting parameters $k = (3.1 \pm 0.03) \times 10^{-2}$ and $L^{Al}_k(0) = (8.23 \pm 0.05) \ \text{pH}$. Despite the likelihood of the aluminum structure being more intricate than a simple inductor, the model agrees very well with the experimental data. The additional dashed lines in Figure \ref{fig:fig2_temp}b represent $\Delta{f_r}(T)$ for increasing values of $L^{Al}_k(0)$, specifically $L^{Al}_k(0) = 15$, $30$, and $50$ pH. This indicates that an increase in $L^{Al}_k(0)$ is associated with a decrease in $\Delta{f_r}$.

\section{Geometrical dependence}

As per Eq. \eqref{eq:Lk_BTdim}, reducing the width of a superconducting wire increases the kinetic inductance, implying a decrease in $\Delta{f_r}$. This observation is consistent with experimental results, where $\Delta{f_r}(T)$ consistently decreases with each successive FIB session, as depicted in Figure \ref{fig:fig2_temp}c. An interesting observation is the emergence of a ``tail" in the resonance frequency following the first and second FIB sessions, indicated by the black arrow in Figure \ref{fig:fig2_temp}c. We attribute this phenomenon to the unavoidable exposure of a part of the Al loop to the ion beam, both while setting up and during the FIB milling. This has an impact on the morphology and properties of aluminum through various physical processes \cite{kohopaa2023wafer}, potentially enhancing the critical temperature \cite{cohen1968superconductivity, deutscher1973transition, levy2019electrodynamics}. This complicates data fitting for both FIB1 and FIB2, as Eq. \eqref{eq:fr_model} does not consider the variation in $T_c$ and the tail in $f_r(T)$. Nevertheless, an evident correlation exists between the increase in $L^{Al}_k(0)$ and the subsequent decrease in $\Delta{f_r}$.

\section{Field dependence}
\label{sec:Bdep}

\begin{figure}[H]
    \centering
    \includegraphics[width=\columnwidth]{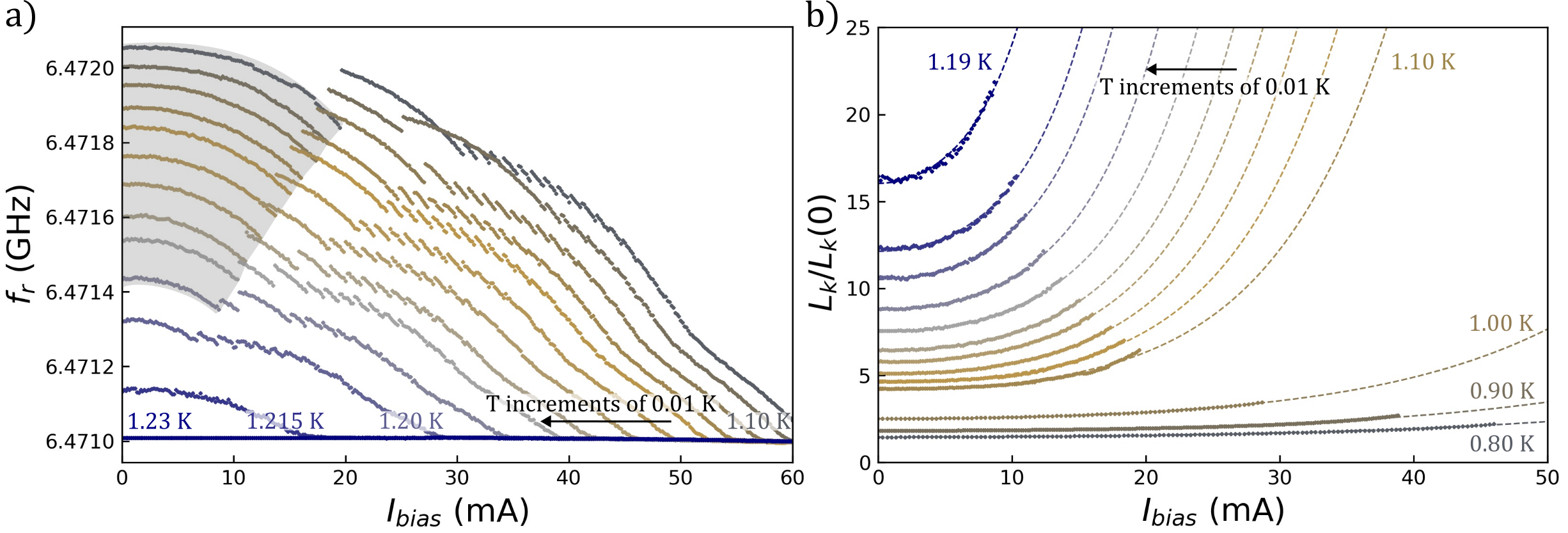}
    \caption{Dependence of resonance frequency and kinetic inductance on bias current. a) The resonance frequency ($f_r$) as a function of the applied bias current ($I_{bias}$) through the external flux bias line as represented in Figure \ref{fig:fig1_sample} for the preFIB sample for different temperatures as indicated. The occurrence of the first phase slip corresponds to the right edge of the shaded grey zone. $I_{bias}$ can be converted to a local magnetic field at the loop as described in the Supporting Information. b) The kinetic inductance fraction $L_k/L_k(0)$ as a function of the applied bias current for the preFIB sample at various temperatures as specified in the legend. The conversion of the $f_r(I_{bias})$ data, presented in panel a) to $L_k(I_{bias})/L_k(0)$ data is detailed in the main text. The dotted lines correspond to fits derived from Eq. \ref{eq:Lkexpand}.}
    \label{fig:fig3}
\end{figure}

Figure \ref{fig:fig3}a depicts the resonance frequency as a function of the bias current $I_{\text{bias}}$ flowing through the flux line at different temperatures. This bias current induces a local magnetic field $B$, which can be estimated using a conversion factor of 38 µT/mA, as outlined in detail in the Supporting Information. The field dependence of the kinetic inductance can be estimated using Eq. \eqref{eq:Lk_BTdim}. As a first-order approximation, the $L_k(T)/L_k(0)$ field-dependence can be compared to the kinetic inductance of a superconducting strip, which can be expanded (for $T < T_c$) in terms of the applied current ($I$) trough the strip according to \cite{zmuidzinas2012superconducting}
\begin{equation}
    \label{eq:Lkexpand}
    L_k(I) = L_k(T,I=0)(1+\frac{I^2}{I_s^2}+ ...) .
\end{equation}

Here, $I_s$ is in the order of magnitude of the critical current, with only even powers considered due to symmetry \cite{zmuidzinas2012superconducting}. While Eq. \eqref{eq:Lkexpand} serves as an estimation for a superconducting strip, it exhibits a close correspondence to the measured system. The bias current induces a local magnetic field to the Al loop, which induces circulating currents, establishing a direct proportionality between the applied bias current and the current $I$ in Eq. \eqref{eq:Lkexpand}.

Above the critical temperature of the Al loop (1.23 K), $f_r(I_{bias})$ remains constant, indicating the Nb resonator is unaffected by the local magnetic field from the bias line, as shown in Figure \ref{fig:fig3}a. However, below $1.23$ K, a distinct impact on the resonance frequency is observed, revealing several different regimes delimited by discontinuities in $f_r(I_{bias})$. Hence, the Nb resonator's frequency variation due to the bias line current depends solely on changes in the Al loop's properties induced by the magnetic field.

From zero bias current to the first discontinuity, $f_r(I_{bias})$ consistently decreases as highlighted by the grey shaded area in Figure \ref{fig:fig3}a. This occurs because as $I_{bias}$ increases, shielding currents rise within the Al loop, reducing $n_s$ and amplifying $L_k$ (Eq. \eqref{eq:Lk_BTdim}). At a certain critical bias current value, a phase slip occurs, locally suppressing the superconducting state and changing the winding number. This reduces the total circulating current in the loop and, consequently $f_r(I_{bias})$ suddenly increases.  

Further information on the system can be gained by focusing on the $f_r(I_{bias})$ regime before the first phase slip. Combining the data for $f_r(T)$ and Eq. \eqref{eq:eq_Lk}, it is possible to extract an $f_r(L_k/L_k(0))$ dependence, assuming the change in $L_k$ is the origin of the temperature-dependent resonance frequency. As such, we can obtain $L_k(T)/L_k(0)$ as a function of the applied bias current, as shown by the solid circles in Figure \ref{fig:fig3}b.

Hence, Eq. \eqref{eq:Lkexpand} up to the 6th-order with fitting parameter $I_s$, can be used to fit the experimental data depicted in Figure \ref{fig:fig3}b, as evidenced by the dashed lines. The proposed protocol demonstrates the possibility to obtain the kinetic inductance fraction of the Al loop, or any inductively coupled structure, through the measurement of the resonance frequency of the coupled superconducting resonator. Unfortunately, Eq. \eqref{eq:Lkexpand} is not applicable for FIB1 and FIB2 since the alteration of only one arm of the loop prevents the entire structure from being treated as a homogeneous superconducting strip. 

As the system's dynamics grow more intricate following the initial entry of flux into the structure, additional distinct regimes emerge. In addition to the multiple small up jumps, 
in the resonance frequency as those shown in Figure \ref{fig:fig3}a, there are also down-jumps of $f_r(I_{bias})$
These events are correlated with a change in the Cooper pair density, possibly induced by a reduction of the circulating current or vortex penetration within the superconducting structure, leading to upward and downward shifts in $f_r(I_{bias})$ respectively. A detailed zoom of these regions is shown in Figure S3. To precisely pinpoint the origin of these regimes, it is necessary to resort to scanning probe techniques with local flux resolution \cite{embon2017imaging,polshyn2018imaging,goa2001real,brisbois2017flux} or simulations in the time-dependent Ginzburg-Landau framework \cite{nulens2022metastable,embon2017imaging,kenawy2020electronically}.

\section{Flux entry}

\begin{figure}[H]
    \centering
    \includegraphics[width=\columnwidth]{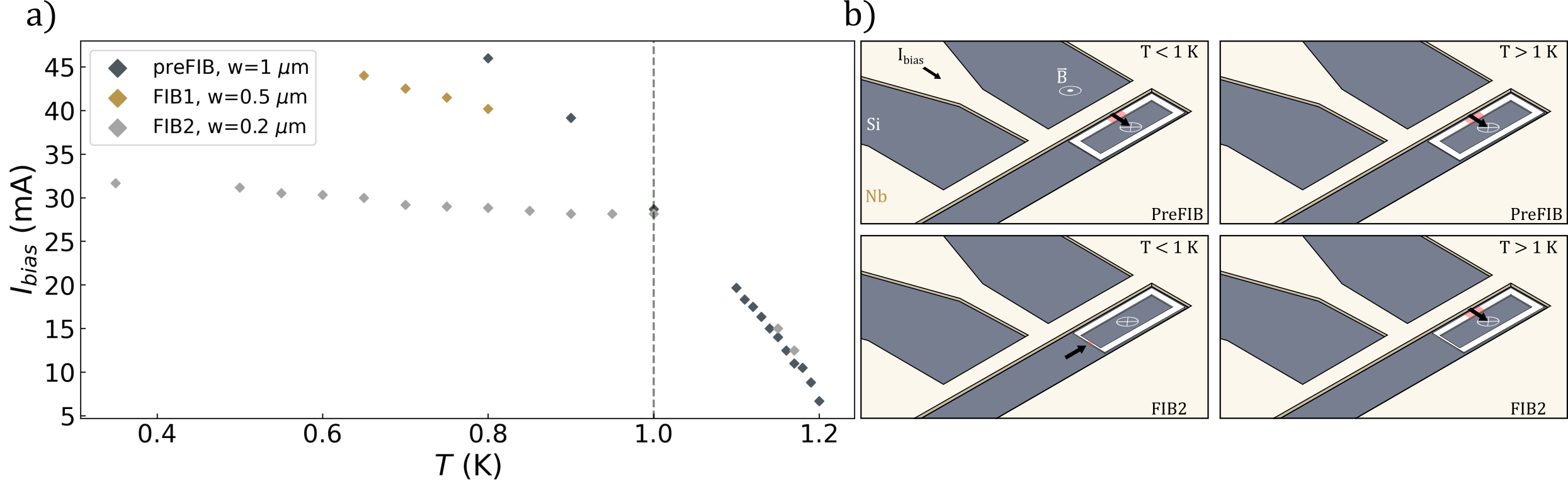}
    \caption{The temperature and geometry dependence of the first phase slip and its schematic representation. a) The bias current ($I_{bias}$) causing the occurrence of the first phase slip, identified by a sudden increase in resonance frequency, for different temperatures and FIB sessions, as illustrated in the legend. b) Schematic representations depict the occurrence of the first phase slip for both the preFIB sample (top row) and the FIB2 sample (bottom row). The left and right panels correspond to regions below and above 1 K, respectively, as indicated by the dashed line in panel a. }
    \label{fig:fig5_BLK_FIB12}
\end{figure}

Similar to the preFIB sample and under identical measurement conditions, $f_r(I_{bias})$ for FIB1 and FIB2 reveal analogous distinct $f_r$ regimes. Narrowing the width of a single section of the Al loop, thereby increasing its kinetic inductance \cite{dausy2021impact}, effectively enhances the probability of phase slip events in that arm of the Al loop. The flux entry point, marked by a discontinuous jump and increase in $f_r$, is depicted as a function of temperature both prior to and following both FIB sessions in Figure \ref{fig:fig5_BLK_FIB12}a.

Notably, the preFIB data points reveal a clear dependence of the initial flux entry on temperature. As the width of the loop is uniform prior to the FIB milling, precisely identifying the location where the flux breaches through the structure is not straightforward. Typically, it occurs at the weakest point in the superconductor. In addition, the magnetic field generated by the flux line is nonuniform, with its maximum value closest to the bias line, as confirmed by COMSOL simulations of the structure in the Supporting Information. Consequently, it is reasonable to assume that the flux enters through the Al arm closest to the bias line as schematically represented in the two top panels of Figure \ref{fig:fig5_BLK_FIB12}b.\\

Measurements from FIB1 below $T < 0.8$~K show a reduced bias current requirement for the first phase slip compared to preFIB, suggesting flux entry via the weak link (500 nm arm) in the Al loop. Above $T > 0.8$~K, vortices penetrate the Al structure before a phase slip occurs, rendering the subsequent $I_{bias}$ value corresponding to a phase slip unrepresentative (see Figure S4). Post-FIB2 measurements showed that below 1 K, flux entry is determined by the weak link, with reduced $I_{bias}$ requirements, confirming FIB1 trends. Above 1 K, data aligns with preFIB samples (Figure \ref{fig:fig5_BLK_FIB12}a), suggesting flux primarily enters through the Al arm closest to the flux bias line, as schematically shown in the bottom two panels of Figure \ref{fig:fig5_BLK_FIB12}b. Moreover, in the pre-FIB, FIB1, and FIB2 cases, the initial frequency jump is observed at a bias current corresponding to a field value significantly surpassing the loop's field periodicity ($\Phi_0/A=17$ µT), indicating a considerable energy barrier between adjacent states \cite{nulens2022metastable,arutyunov2012quantum}.

\section{Conclusion}
\label{sec:concl}

We present a novel approach to probe superconducting nanostructures coupled to a superconducting resonator via high-frequency spectroscopy. Utilizing a flux bias line to induce a localized magnetic field, we investigate the superconducting properties of an Al loop without affecting the Nb readout resonator. This method is readily applicable to superconducting structures with critical temperatures below 4 K, as the readout resonator's superconducting properties remain unchanged below this temperature. However, this limitation could be addressed by employing a readout resonator with a higher $T_c$, such as NbTiN \cite{nulens2023catastrophic} or YBCO \cite{valenzuela1989high,arzeo2014microwave}.

The readout resonator's resonance frequency is directly affected by the kinetic inductance of the Al loop, exhibiting distinct dependencies on temperature and magnetic field. Field dependency unveils specific regimes associated with flux-induced changes in the loop's Cooper pair density. Our focus lies on the region preceding the initial discontinuous jump in $f_r$. In this case, circulating currents intensify to shield the increasing magnetic field, reducing the Cooper pair density, and subsequently lowering the resonance frequency until the first phase slip occurs. This phase slip diminishes the total circulating current leading to a sudden increase in $f_r$. The good agreement between the fit following Eq.\eqref{eq:Lkexpand} and the first region in the experimental data demonstrate that we can effectively probe the change in the kinetic inductance without affecting the superconducting state.

To advance our understanding of the phase slip dynamics and its kinetic inductance dependence, we utilize FIB milling to modify one arm of the Al loop, creating a weak link that facilitates the formation of phase slips and shifts the initial flux entry point. While reducing the width of one arm to 200 nm accelerates the initial flux entry point, an energy barrier persists, preventing phase slips at half-flux-quantum typically associated with a quantum phase slip regime. Considering the dimensions required to observe this effect \cite{astafiev2012coherent,peltonen2013coherent,skacel2019development}, this limitation is not surprising. Nonetheless, these measurements show the feasibility of our proof-of-principle device as a platform for phase slip manipulation and detection, keeping in mind that using e-beam lithography for the weak link fabrication, alternative material choices, and reducing the film thickness are all strategies that can be used to decrease the energy barrier between adjacent energy states.

\section{Supporting Information}
Additional device fabrication details, FIB milling approach, current to field conversion and corresponding COMSOL simulations, experimental data of the $f_r(I_{bias})$ dependence for FIB1 and FIB2 (PDF).

\section{Author Contributions}
L.N. and B.R. designed the devices, L.N. and K.B. fabricated the devices, J.S. executed the FIB milling, L.N. and O.J.Y.H. executed the measurements, N.L and A.V.S. executed Comsol simulations. L.N., D.A.D.C., and J.V.V. took the lead in writing the manuscript with the help of all other authors. All authors discussed the results and commented on the manuscript. J.V.V. and M.J.V.B. initiated and supervised the research.

\section{Acknowledgements}
The authors would like to thank Claudia Fleischmann and André Vantomme for providing access to the FIB-SEM (supported by FWO-Hercules through project AKUL 15-22 ZKD 1037) and Christian Haffner for his support regarding the experimental development of the high-frequency setup. This work is supported by Research Foundation Flanders (FWO) grant number 11K6523N, the EU COST action SUPERQUMAP CA21144, the Fonds de la Recherche Scientifique - FNRS under the grant Weave-PDR T.0208.23 and has received funding from the BOF IBOF/23/65 project. N.L. acknowledges support from FRS-FNRS (Research Fellowships FRIA). \\

\clearpage

\bibliography{Ref.bib}
\centering
For Table of Contents Only
\begin{figure}[H]
    \centering
    \includegraphics[width=0.5\textwidth]{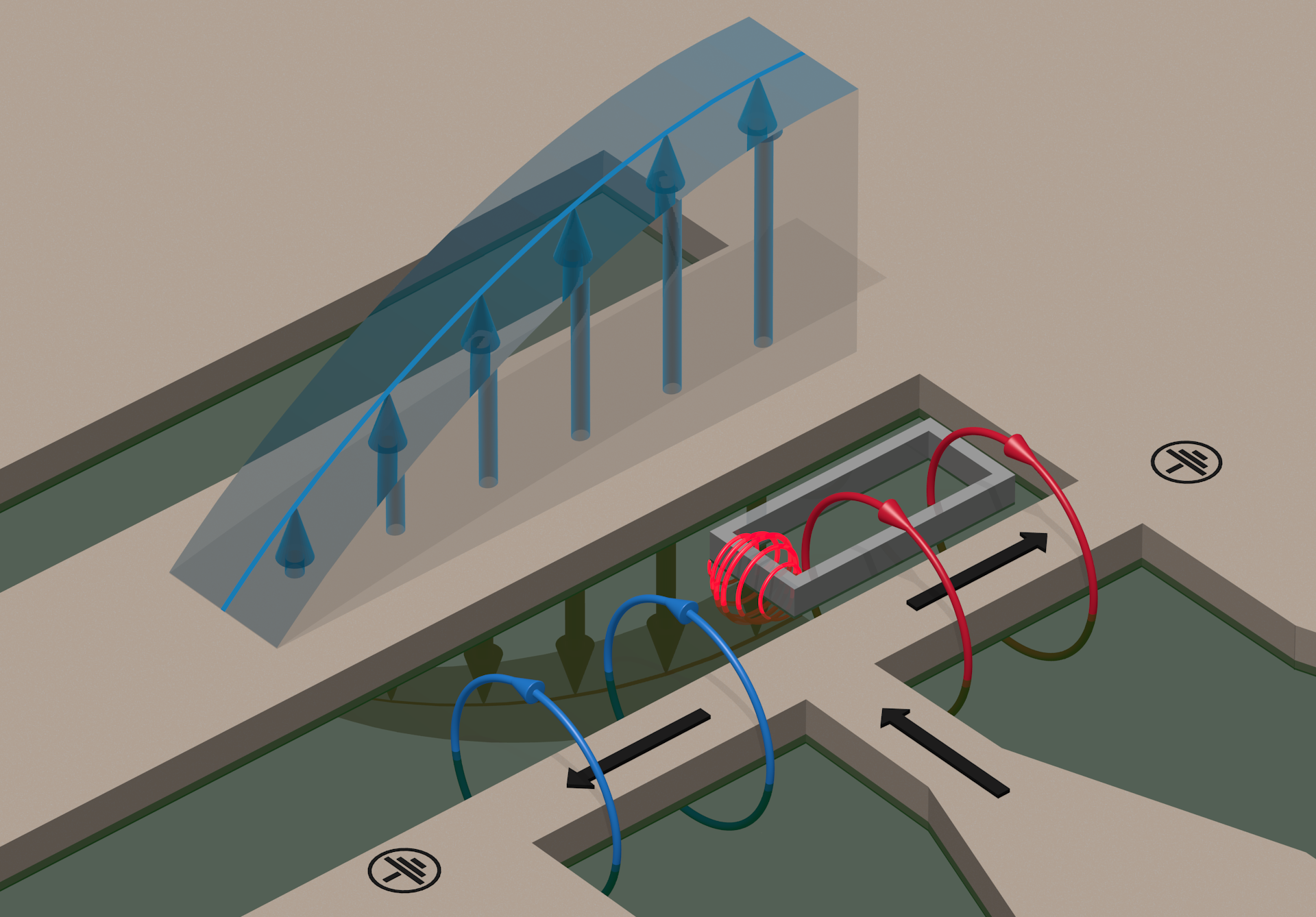}
\end{figure}

\end{document}

\section{Experimental Section}
\subsection{Sample fabrication}
The Si substrate, measuring $(2 \times 2)$ cm$^2$, underwent immersion in a $10\%$ hydrofluoric acid solution for 60 seconds before being transferred to a high-vacuum environment (\textless $10^{-6}$ mbar). Subsequently, molecular beam epitaxy (MBE) was conducted, depositing a 100 nm thick niobium film onto both the substrate and a reference substrate at a rate of 1.2-1.3 \AA/s under room temperature conditions, while maintaining a pressure of approximately $10^{-9}$ mbar. SQUID magnetometry analysis of the deposited Nb reference layer revealed a superconducting critical temperature ($T_c$) of approximately 9.2 K. After applying a PMMA-5C resist layer to the sample, multiple devices were patterned using standard e-beam lithography (in a customized
electron beam lithography platform from Raith GmbH). Subsequently, a chlorine-based etching process was employed, followed by a dicing procedure to shape the samples into ($7 \times 3$) mm$^2$ pieces. One of these devices was wire-bonded and measured, with results detailed in the main text. After the wirebond removal, a resist stack of MMA/PMMA was spin-coated onto the sample, followed by e-beam lithography to pattern the loop structure. Afterward, a 50 nm aluminum deposition took place through MBE at room temperature, with a deposition rate of 1.2 \AA/s and pressure approximately $10^{-9}$ mbar. Following this, excess aluminum was removed using a lift-off procedure in acetone.

\subsection{Focussed Ion Beam milling}
The modification of the Al loop was carried out using the Helios G3 CX dual beam FIB-SEM from Thermofischer Scientific. A $^{69}$Ga ion beam with a beam energy of 30 kV and a current of 80 pA was employed to reduce the size of the loop’s arm. For this, a rectangular milling pattern of appropriate size was placed to partially overlap the targeted arm. The beam impinged the sample at an angle of 52\textdegree \text{ }with respect to the sample's surface normal. While the primary goal of FIB milling is to locally reduce the width of one arm, it is important to consider unavoidable ion beam exposure of a part of the aluminum loop, both while setting up and executing the patterned ion beam milling. When the Ga ion beam hits the (nano)crystalline Al surface, a combination of sputtering, Ga-implantation, and amorphization occurs. Subsequently, the sample surface can (re-)oxidize and potentially recrystallize (by thermal annealing) in ambient conditions. Such a change in the film's morphology has an impact on its properties, for example an alteration of the critical temperature of (a part of) the Al structure \cite{kohopaa2023wafer}.

\section{Supplemental note}
\subsection{Current-field conversion}
\renewcommand{\thefigure}{SN\arabic{figure}}

\begin{figure}[H]
    \centering
    \includegraphics[width=0.6\textwidth]{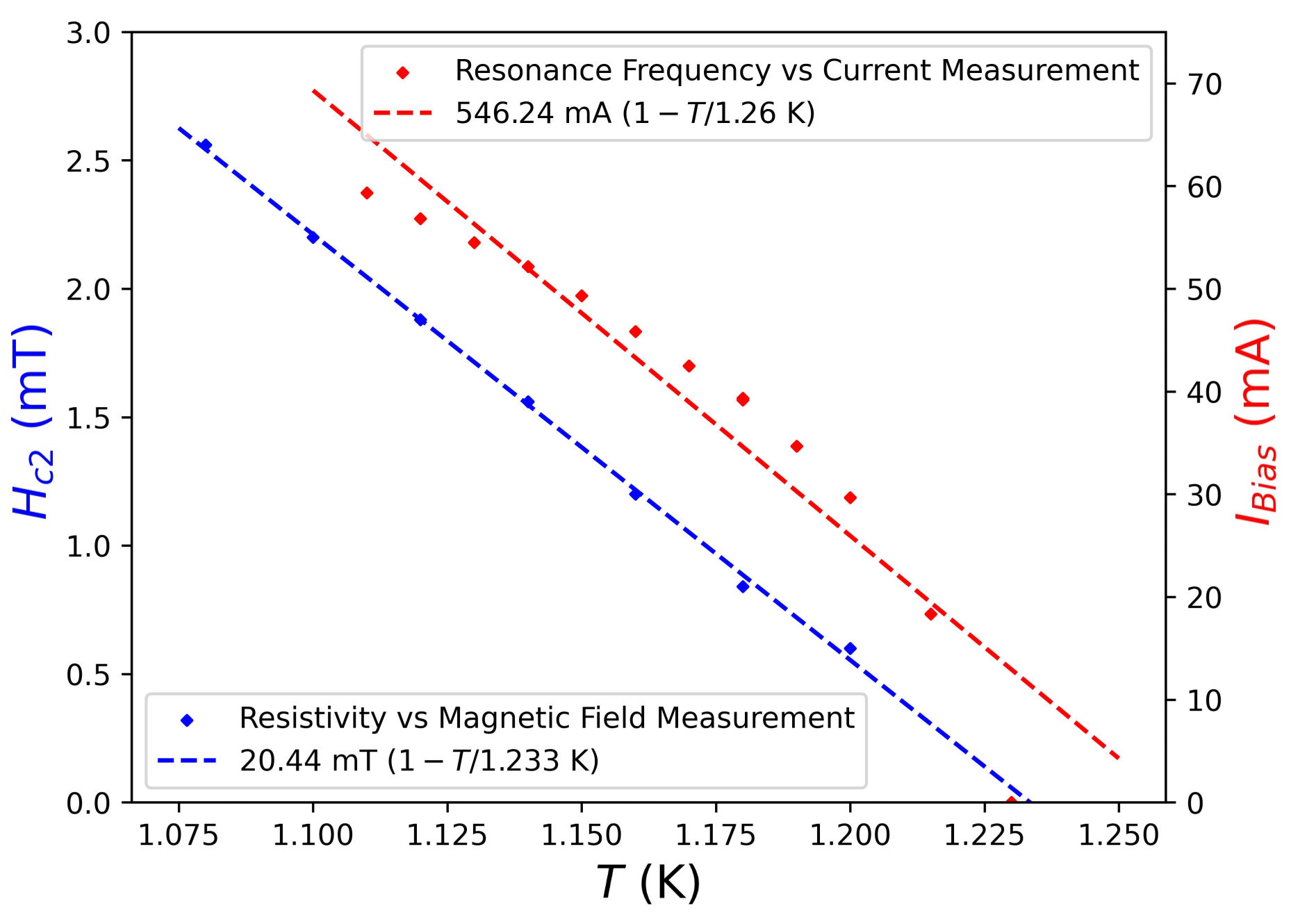}
    \caption{Conversion from bias current to local magnetic field. The blue data points represent the upper critical field ($H_{c2}$) of a reference aluminum bridge. These points are obtained with an applied current of 1 µA as a function of temperature, with $H_{c2}$ determined as 90$\%$ of the normal resistance ($R_n$). The blue dashed line corresponds to a linear fit of these data points, and the resulting expression is provided in the legend. The red data points are acquired from $f_r(I_{bias})$ measurements, where the dot marks the location where the resonance frequency converges with the $f_r(I_{bias})$ curve at 1.23 K. The red dashed line represents a linear fit of these data points, and the resulting expression is indicated in the legend.}
    \label{fig_sup_1_IB}
    \label{fig:fig1_sample}
\end{figure}

The bias current passing through the flux line can be converted into an estimated local magnetic field. This comparison is established by examining the temperature-dependent upper critical field ($H_{c2}$) of a reference aluminum bridge (with dimensions: width = 1 µm, thickness = 50 nm, and length = 120 µm) against the temperature-dependent bias current $I_{c-bias}$ required for the transition of the aluminum loop to its normal state, as depicted in Figure 3a. In Figure \ref{fig_sup_1_IB}, the blue data points represent $H_{c2}(T)$, with the blue dashed line depicting a linear fit of the data: $H_{c2}= 20.44 \text{mT}(1-T/1.233 \text{K})$. The red data points correspond to $I_{c-bias}(T)$, with a linear fit shown by the red dashed lines: $I_{c-bias}= 546.24\text{mA} (1-T/1.26\text{K})$. From these linear fits, a ratio of $H_{c2}(T)/I_{c-bias}(T) \approx 38$ µT/mA is obtained.

\begin{figure}[H]
    \centering
    \includegraphics[width=0.6\textwidth]{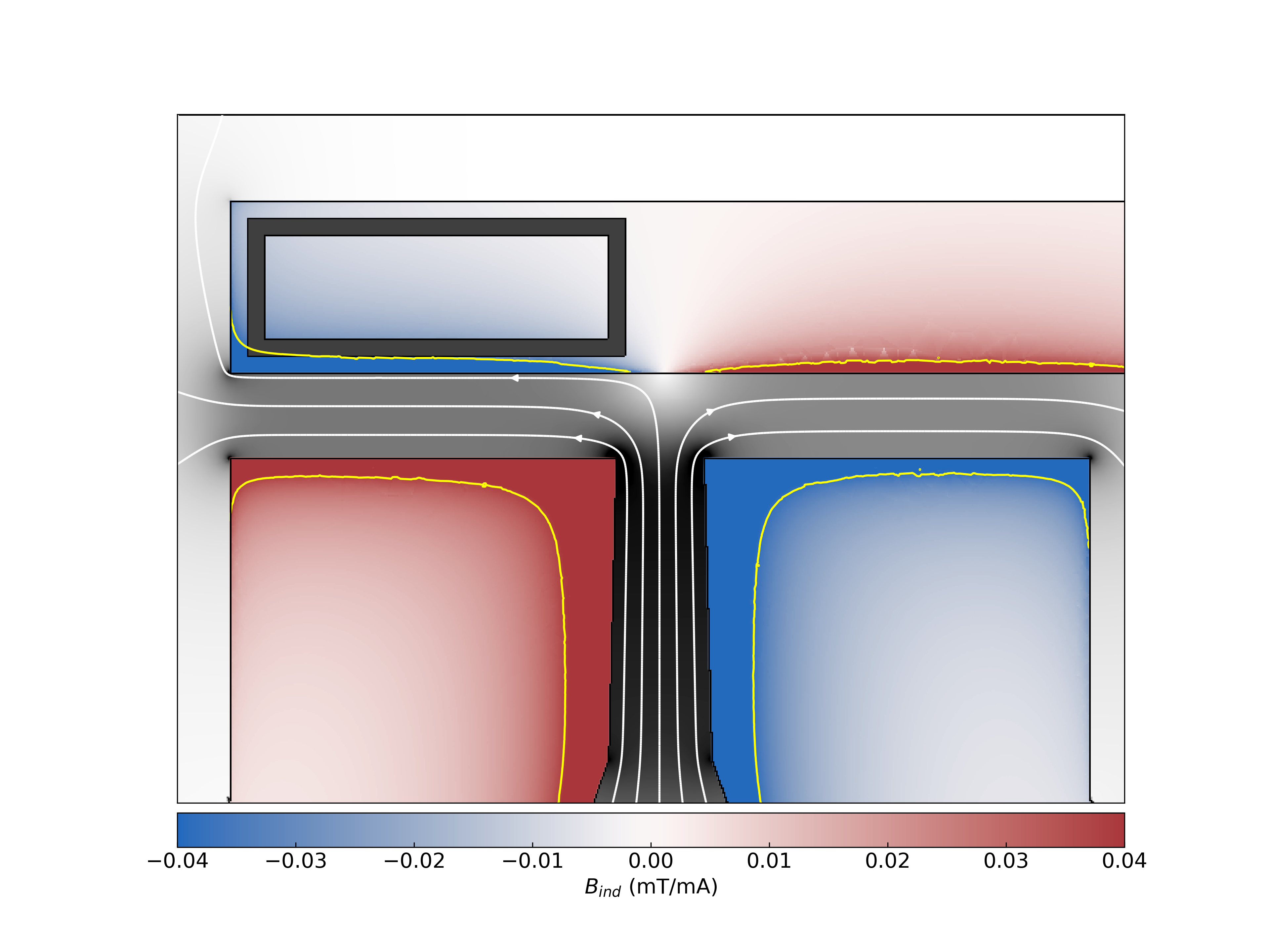}
    \caption{Simulated out-of-plane magnetic field generated by the current distribution in the Nb flux-bias. The Al loop position is depicted in gray. The frontier delimiting the region where the magnetic field is sufficient to suppress the loop’s superconductivity is shown in yellow and corresponds to a value of $B_{ind}$ = 38 µT/mA. This line intersects the loop on the side closer to the flux line, where the flux is expected to penetrate into the loop close to $T_c$.}
    \label{fig_SN2}
\end{figure}

The magnetic field intensity generated by the current fed into the flux-bias line, was simulated using successively the \textit{Electric Currents} and \textit{Magnetic Fields} modules of the COMSOL Multiphysics software. The sample layout was reproduced to first simulate the current density distribution in the bias line, assuming it as a normal metal. Based on this result, the out-of-plane magnetic field distribution in the Nb gap regions was computed. Both the current stream lines and the magnetic intensity map are combined in Figure \ref{fig_SN2} where the current density is shown in inverted grayscale onto the Nb parts whereas the magnetic field landscape is shown in the gaps. From the field calibration presented in Figure \ref{fig_sup_1_IB} we determined the ratio $H_{c2}(T)/I_{c-bias}(T)$ at which the Al loop switches to the normal state. Using the numerical simulations, we are able to identify the delimiting region where this ratio $H_{c2}(T)/I_{c-bias}(T)$ takes place, or in other words, the physical region where the local field exceeds the critical field $H_{c2}(T)$ of the Al loop. This line, which is indicated with yellow color in Figure \ref{fig_SN2}, intersects the Al loop on the side through which the flux quantum enters during a phase slip event at temperatures close to $T_c$ (see Figure 4(b)). Note that since the simulation was performed assuming a normal metal behavior, the current distribution is assumed uniform in straight portions whereas in superconductors in the Meissner state, the current flows preferentially closer to the borders of the conductor strip, which further increases the magnetic field magnitude close to the border of the superconductor \cite{benkraouda1998critical}, moving the threshold frontier deeper in the Al loop.

\section{Supporting Figures}
\renewcommand{\thefigure}{S\arabic{figure}}

\begin{figure}[H]
    \label{fig_sup_3_frB_FIB_zoom}
    \centering
    \includegraphics[width=\textwidth]{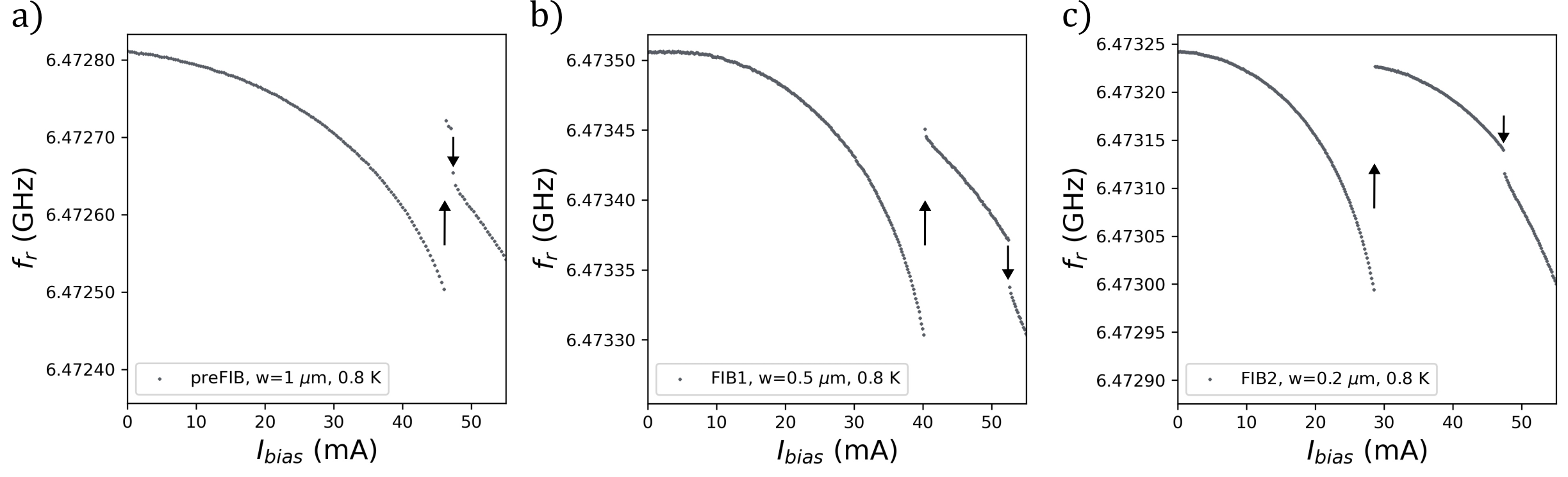}
    \caption{A zoom of the resonance frequency as a function of the local magnetic field arising from the current $I_{bias}$ through the bias line at 0.8 K for the a) preFIB, b) FIB1 and c) FIB2 device. The upward and downward arrows correspond with upward and downward jumps of $f_r(I_{Bias})$ as described in the main text, respectively.}
    \label{fig:fig1_sample}
\end{figure}

\begin{figure}[H]
    \label{fig_sup_2_frB_FIB}
    \centering
    \includegraphics[width=\textwidth]{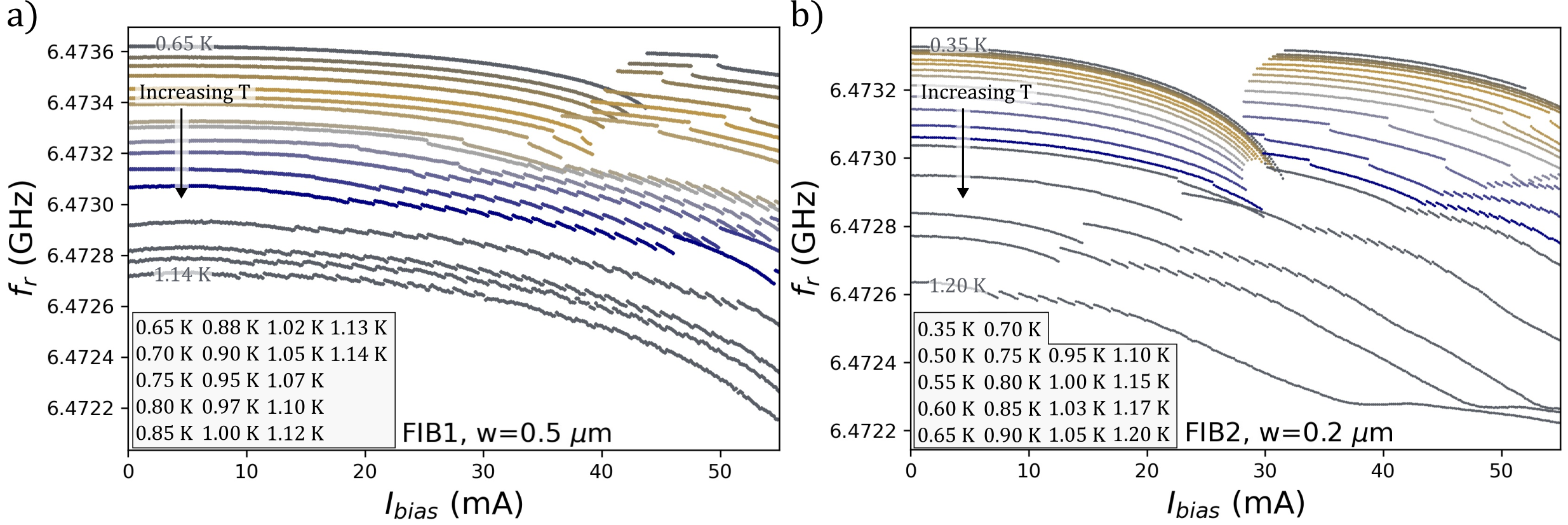}
    \caption{The resonance frequency ($f_r$) as a function of the applied bias current $I_{bias}$, which gives rise to a local magnetic field at the loop. Panel a) depicts the results for FIB1, with temperature values specified in the legend, while panel b) presents the findings for FIB2.}
    \label{fig:fig1_sample}
\end{figure}

\bibliography{Ref_sup.bib}